# Structure of nanoparticles embedded in micellar polycrystals

Elisa Tamborini[1,2], Neda Ghofraniha[1,2], Julian Oberdisse[1,2,3], Luca Cipelletti[1,2] and Laurence Ramos[1,2] *

[1] *Université Montpellier 2, Laboratoire Charles Coulomb UMR 5221, F-34095, Montpellier, France*
[2] *CNRS, Laboratoire Charles Coulomb UMR 5221, F-34095, Montpellier, France*
[3] *Laboratoire Léon Brillouin, UMR12, CEA/CNRS CEA Saclay 91191 Gif-sur-Yvette Cedex, France*

**Abstract**

We investigate by scattering techniques the structure of water-based soft composite materials comprising a crystal made of Pluronic block-copolymer micelles arranged in a face-centered cubic lattice and a small amount (at most 2% by volume) of silica nanoparticles, of size comparable to that of the micelles. The copolymer is thermosensitive: it is hydrophilic and fully dissolved in water at low temperature ($T \sim 0°C$), and self-assembles into micelles at room temperature, where the block-copolymer is amphiphilic. We use contrast matching small-angle neuron scattering experiments to probe independently the structure of the nanoparticles and that of the polymer. We find that the nanoparticles do not perturb the crystalline order. In addition, a structure peak is measured for the silica nanoparticles dispersed in the polycrystalline samples. This implies that the samples are spatially heterogeneous and comprise, without macroscopic phase separation, silica-poor and silica-rich regions. We show that the nanoparticle concentration in the silica-rich regions is about tenfold the average concentration. These regions are grain boundaries between crystallites, where nanoparticles concentrate, as shown by static light scattering and by light microscopy imaging of the samples. We show that the temperature rate at which the sample is prepared strongly influence the segregation of the nanoparticles in the grain-boundaries.

# 1. Introduction

Huge efforts are currently devoted to the design of new materials where remarkable mechanical, optical, or electric and magnetic properties are achieved by assembling nanoscopic constituents [1]. In this framework, the confinement of nanoparticles (NPs) in a background matrix is considered as one of the most promising techniques with many applications in photonics [2], bio-engineering [3], and electronics [4]. Strategies for assembling nanoparticles in a macromolecular matrix include "active" approaches based on the application of external fields [5], using thermal [6], electric [5] or capillary [7] forces, as well as "passive" methods relying on self-assembly in a macromolecular matrix [8, 9], or at the interface of phase-separating fluid mixtures [10]. In many of these new materials, a crucial step is the control at the microscopic scale of the dispersion of the NPs, since their spatial distribution can influence significantly the material properties.

Block copolymers and surfactant phases are a popular choice as a background matrix, because under appropriate conditions they organize themselves in spatially ordered mesophases, such as lamellae, cylinders arranged on a triangular lattice (hexagonal phases) or spheres arranged, e.g., on a cubic lattice. One then exploits the spontaneous microscopic organization of the matrix to impart the desired order to the NPs. An efficient way to implement this concept is to use nanoparticles that interact preferentially with one of the groups of the surfactant molecules or one of the blocks of the copolymer [11, 12], leading to their microscopic confinement in the domains rich in that group or block. A recent example is provided by Sharma *et al.* [13, 14], who studied a composite of silica particles in a water-based hexagonal mesophase of a nonionic surfactant. Their work illustrates nicely the role of particle size on confinement: hydrophilic particles smaller than the cylinder spacing were efficiently confined in-between adjacent cylinders, while particles slightly larger than the cylinder spacing were partially expelled in the grain boundaries between ordered cylinder domains. Particles even larger were fully segregated in the grain boundaries, forming structures up to tens of microns in size. Segregation of NPs in the defects of a crystalline matrix has also been observed in two-dimensional systems, e.g. in particle-doped thin films of diblock copolymers [15, 16]. NPs have also been found to be expelled from a self-assembled lipid-based matrix [17]

Similar results were also obtained by Lin and coworkers [18], who showed that hydrophilic-functionalized silica and gold particles could be incorporated very efficiently (up to about 35% by volume) in block-copolymers phases with various morphologies (lamellar, hexagonal, cubic), provided that the particles were sufficiently small to fit in the hydrophilic microdomains of the melt. Interestingly, the addition of nanoparticles modified the matrix, inducing microphase separation in melts otherwise disordered, enhancing order in mesophases that were already microscopically structured, or changing the crystalline order from face-centered to body-centered cubic. The interplay between the host phase and the inclusions can be even more complex when the NPs interact strongly with one or more constituents of the matrix. Sun and Raghavan [19] have studied the effect of the addition of Laponite nanoplatelets to aqueous solutions of triblock copolymers of the Pluronics family. Initially, the copolymer strongly adsorbs on the platelets, resulting in a fluid suspension of copolymer-coated platelets. At higher copolymer concentrations, yet smaller than those required for the copolymer solution alone to solidify, the material undergoes a fluid-solid transition, driven by the formation of a disordered gel due to the depletion forces exerted on the platelets by copolymer micelles. Boucenna *et al.* have explored a similar system, but in a more concentrated regime, were the copolymer would normally form a crystal of micelles, due to entropic reasons [20]. They find that the addition of Laponite melts the crystalline mesophase and attribute this effect to the adsorption of the copolymer on the platelets, which decreases the amount of copolymer available for micelle formation, thereby lowering the micelle concentration below the freezing volume fraction.

In most of the examples cited above, the microdomains of the matrix and the NPs differ not only by chemical composition, but also in shape, as for the spherical inclusions in lamellar or cylindrical phases of Refs. [11, 12, 13, 14, 18] or the platelets immersed in a crystal of spherical micelles of Ref. [20]. In a distinct yet related line of research, scientists in the colloid community have explored the effect of the addition of spherical particles to a colloidal crystal made of particles of the same shape and composition. While inclusions that are significantly larger than the host particles induce local defects but do not modify otherwise the crystalline order [21], addition of particles with comparable size can lead to the formation of colloidal alloys where both particle species participate to a new, more complex crystal unit cell [22, 23].

In the present paper, we use neutron and light scattering to investigate the local order and the mesoscopic structure of a novel composite material comprising spherical Pluronics micelles and small amounts (up to 2%) of spherical silica nanoparticles of comparable size. Using neutron scattering contrast variation methods, we are able to probe separately the microscopic micellar arrangement and that of the host particles, as a function of NP content. Moreover, we take advantage of the temperature dependence of the micelle volume fraction in Pluronics system to vary the rate at which the matrix is brought from the fluid to the crystalline phase, in order to explore systematically the dependence of the composite structure on the crystallization rate. We find that the crystalline structure of the matrix is remarkably robust with respect to changes in the crystallization rate and NP concentration. By contrast, both control parameters influence markedly the spatial distribution of the NP, which is found to be non-uniform. Indeed, neutron scattering data indicate that silica particles accumulate in silica-rich regions. Light scattering and optical microscopy reveal that these regions correspond to a three-dimensional network of grain boundaries. As previously modeled for a related experimental system [24], we find that the characteristic size of the network depends on the crystallization rate and NP content.

## 2. Materials and Methods

### 2.1. Materials and sample preparation

The copolymer samples are composed of an aqueous suspension of Pluronic F108, a commercial PEO-PPO-PEO triblock copolymer purchased from Serva Electrophoresis GmbH, of molecular weight 14000 g/mol, where PEO and PPO denote polyethylene oxide and polypropylene oxide, respectively. On average, each PEO block is made of 132 monomers, and the central PPO block is made of 52 monomers. The co-polymer is fully dissolved at $T \cong 0°C$; upon heating, the PPO central block becomes increasingly hydrophobic, resulting in the formation of micelles with a diameter of 22 nm [25], whose number increases with $T$, eventually leading to a crystalline phase at room temperature due to micelle crowding [26, 27], as observed also in similar copolymer systems [28, 29]. We add to the copolymer solution small amounts (up to 2% v/v) of Bindzil® plain silica colloidal nanoparticles (kind gift from Eka Chemicals), with an average diameter $d = 30$ nm and a relative polydispersity of 19 % (sample type 40/130), as determined by transmission electron microscopy. To check for the adsorption of F108 on silica, additional experiments have also been performed with smaller silica NPs (Bindzil® plain silica, sample type 30/220, with $d = 19$ nm and relative polydispersity 26%). In all samples, the copolymer concentration is 34% w/w, regardless of the amount of NPs.

Samples are prepared by adding the F108 flakes to a suspension of nanoparticles. As a solvent, we use mixtures of $H_2O$ and $D_2O$, as detailed in Secs. 2.2 and 2.3. The solvent-NPs-polymer dispersions are thoroughly stirred and kept at 3°C for a few days to obtain fully homogeneous fluid suspensions. The volume fraction of silica NPs, $\phi$, is varied between 0 and 2%. In most cases, the pH of the samples was not adjusted (it is about 6 without NPs, 9 with NPs). In a few samples, we have adjusted the pH of samples containing NPs to 7 and 5 by adding sulfuric acid reaching a molar concentration of 1.2 mM and 2.4 mM in water, respectively, in order to test the effect of pH on our system. Prior to measurements, the samples, which are always kept at 3°C, are introduced in the sample chamber (Hellma quartz cells with optical path 1 or 2 mm for neutron scattering experiments, optical glass cells with optical path 2 mm, respectively 1 mm, for small, respectively ultra-small angle static light scattering, borosilicate cylindrical cells with inner diameter of 8 mm for static light scattering, and microscopy slides for light microscopy visualization). Two procedures are used subsequently to bring the samples from the fluid phase to the crystalline phase. The first procedure consists in placing the sample chambers in a copper container immersed in a Haake thermal bath, whose temperature is raised from 3 °C to 20 °C at a controlled temperature rate, $\dot{T}$. Three temperature rates, 0.02 °C/min, 0.007 °C/min, and 0.001 °C/min, are used. Alternatively, the sample chambers are immersed directly in a water bath at 21°C, to achieve a faster heating rate. In this case, we have determined that the temperature of the sample rises rapidly initially, and then slows down. In the relevant temperature range where crystallization occurs (15-18°C, as determined independently by calorimetry and rheology), the heating rate is about 2 °C/min. Once crystallized, the samples are kept at room temperature until measurements are performed.

## 2.2. Small Angle Neutron Scattering

Small-angle neutron scattering (SANS) experiments have been performed on the beamline PACE at the Laboratoire Leon Brillouin, Saclay, France. To cover a broad range of scattering vectors $q$ (0.003 Å$^{-1}$ $\leq q \leq$ 0.4 Å$^{-1}$), three experimental configurations are used, $D$ = 4.5 m, $\lambda$ = 12 Å; $D$ = 4.5 m, $\lambda$ = 6 Å; $D$ = 1 m, $\lambda$ = 6 Å, where $D$ is the sample-detector distance and $\lambda$ is the neutron wavelength. Here, $q = 4\pi\lambda^{-1}\sin(\theta/2)$ is the magnitude of the scattering vector and $\theta$ is the scattering angle. The instrumental resolution is $\Delta q/q = \Delta\lambda/\lambda$ = 10% (FWHM). Empty cell background subtraction, calibration by light water in a 1 mm-thick quartz cell, and absolute determination of scattering cross-sections, $I(q)$, per sample volume in cm$^{-1}$ were performed using standard procedures based on an independent measurement of the incoming beam intensity [30]. The incoherent background was estimated from the known high-$q$ Porod law ($I \sim q^{-4}$) of silica particles or the $q^{-2}$ decrease of the scattered intensity at high $q$ of the polymer micelles [30]. Solvent contrast variation using mixtures of H$_2$O (scattering length density $\rho_{H_2O}$ = -0.56 x 10$^{10}$ cm$^{-2}$) and D$_2$O (scattering length density $\rho_{D_2O}$ = 6.34 x 10$^{10}$ cm$^{-2}$) was used to extinguish either the signal from the silica NPs (by matching the solvent scattering length density to that of silica, $\rho_{sil}$ = 3.47 x 10$^{10}$ cm$^{-2}$) or that from the polymer (by matching the solvent scattering length density to that of the polymer, $\rho_{pol}$ = 0.41 x 10$^{10}$ cm$^{-2}$ as estimated from the scattering densities of PEO and PPO). The required solvent compositions are 39% H$_2$O / 61% D$_2$O w/w to match the silica NPs and 85% H$_2$O /15% D$_2$O w/w to match the copolymer. For the samples prepared without nanoparticles, we use pure D$_2$O as a solvent, to maximize the contrast between the copolymer and the solvent and to reduce incoherent scattering from hydrogenated species. For all measurements, temperature is fixed at 22°C.

## 2.3. Static light scattering, small-angle and ultra-small angle light scattering

Light scattering measurements are performed in order to extend the range of $q$ vectors in the low $q$ range, covering almost three additional decades in scattering vector, $4.0 \times 10^{-6} \leq q \leq 3.2 \times 10^{-3}$ Å$^{-1}$. Such a large range in $q$ is obtained by combining static light scattering (SLS), small angle light scattering (SALS) and ultra-small angle light scattering measurements (USALS). For all light scattering experiments, the solvent is H$_2$O, with a refractive index $n$ = 1.33 (the refractive index of the copolymer is 1.38). The magnitude of the scattering vector is calculated from the scattering angle according to $q = 4\pi n \lambda_0^{-1} \sin(\theta/2)$, where $\lambda_0$ is the in-vacuo wave length of the laser source. Static light scattering measurements cover the range $5.66 \times 10^{-4} \leq q \leq 3.19 \times 10^{-3}$ Å$^{-1}$ ($20 \deg \leq \theta \leq 157 \deg$). They have been performed using a standard apparatus comprising an Argon ion laser operating at $\lambda_0$ = 514 nm, an Amtec goniometer and a Brookhaven BT9000 correlator. Data are averaged over 16 runs and the sample cell is slightly rotated in between runs, to insure adequate averaging. The SALS apparatus is custom-made and will be described in detail elsewhere. It covers the range $1.2 \times 10^{-5} \leq q \leq 5.6 \times 10^{-4}$ Å$^{-1}$ ($0.5 \deg \leq \theta \leq 24.5 \deg$), using a HeNe laser as a source ($\lambda_0$ = 633 nm) and a CCD camera as a detector. Data are averaged azimuthally, i.e. over different orientations of the scattering vector, keeping constant its magnitude. The contribution of stray light due to imperfections in the apparatus lenses and cell walls is corrected for following the procedure described in [31]. The USALS setup is similar to that described in [32]; it covers the range $4.0 \times 10^{-6} \leq q \leq 4.9 \times 10^{-5}$ Å$^{-1}$ ($0.2 \deg \leq \theta \leq 2.13 \deg$). The setup is equipped with the same source and detector as the SALS apparatus and data are processed in the same way.

**2.4. Light microscopy**

The sample chambers are made by two coverslips separated by a 250 μm thick 16x16 mm$^2$ double-adhesive gene frame (Thermo Scientific). Imaging is performed using an upright Leica microscope with an air x63 objective or with an oil x100 objective. Bright field and differential interference contrast (DIC) are used.

# 3. Results

## 3.1. Structure of the suspension of micelles

The crystalline structure of the suspensions of polymeric micelles in $D_2O$ and without silica nanoparticles has been characterized by SANS. The intensity spectra for samples prepared with different temperature rates, from 0.001 °C/min to 2 °C/min, are shown in Figure 1. All the scattering curves nearly overlap, indicating that the polymer structure on the length scale probed by the neutrons is not affected by the choice of the heating rate. Minor differences between the curves, in particular in the low-angle limit, as well as in the exact peak position, can be noted.

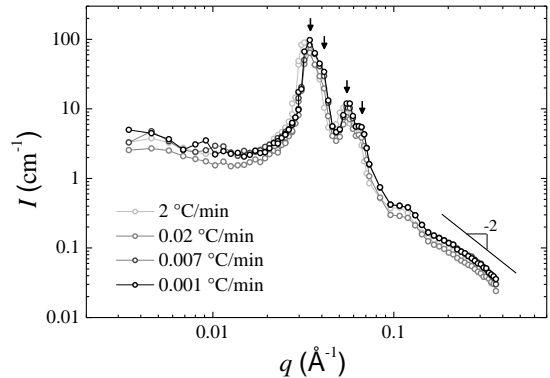

**Fig 1:** Scattering intensity versus wave vector for samples without nanoparticles, prepared using different temperature rates, as indicated. The arrows point to the four main diffraction peaks, whose positions are in the ratio $1:\sqrt{4/3}:\sqrt{8/3}:\sqrt{11/3}$.

Four diffraction peaks are measured. The position of the first order peak is $q_1 = (0.0341 \pm 0.0008)$ Å$^{-1}$ and the positions of the three other peaks relative to the first peak are compatible with the values $\sqrt{4/3}$, $\sqrt{8/3}$, and $\sqrt{11/3}$, respectively (see the arrows in Fig. 1). This indicates that the micelles are organized on a face-centered cubic (fcc) crystal lattice, with a lattice parameter $a = (319 \pm 7)$ Å. We take for the error bars the standard deviation of the measurements of eight different samples (Figs. 1 and 2). Note that the first and second diffraction peaks are not well resolved due to the limited instrumental resolution, as already observed for a similar micellar cubic phase [33].

In order to probe the structure of the micelles in the presence of nanoparticles, contrast-matching SANS experiments are performed on samples prepared with different amounts of silica NPs in a $H_2O/D_2O$ mixture, which index-matches the NPs scattering. The only contribution to the coherent scattering intensity then stems from the micelles. Figure 2 displays the scattered intensity for silica-matched samples containing various amounts of silica ($\phi$=0.1%, 0.5%, 1% and 2%) and for a sample without nanoparticles, all samples being prepared with a heating rate of 0.007 °C/min. The scattering curves are plotted in a normalized way, $I/\Delta\rho^2$, where $\Delta\rho$ is the contrast between the polymer and the solvent. This normalization allows one to directly compare on the same scale the silica-containing samples (where the solvent is a mixture $H_2O$ and $D_2O$) and a sample without nanoparticles (where the solvent is $D_2O$). At intermediate and high $q$-values ($q>0.025$ Å$^{-1}$), the intensities overlap nicely, thus indicating that the crystalline fcc order of the micelles is robust and unaffected by the addition of moderate amounts (up to 2%) of silica nanoparticles.

The ratio of the peak intensities of the first and third peak is comparable to that reported in the literature for similar micellar cubic phases [33, 34]. After correction for the micelle form factor as measured independently in a dilute polymer solution at 2 %, this ratio is consistent with a Debye-Waller factor, $\exp(-q^2u^2/3)$, with $u = 42$ Å. Here $u$ characterizes the typical deviation of the centers of mass of the micelles from the exact lattice positions. Importantly, we find that the value of $u$ is constant and independent of NP concentration, showing that the NPs do not perturb the crystalline order of the micelles.

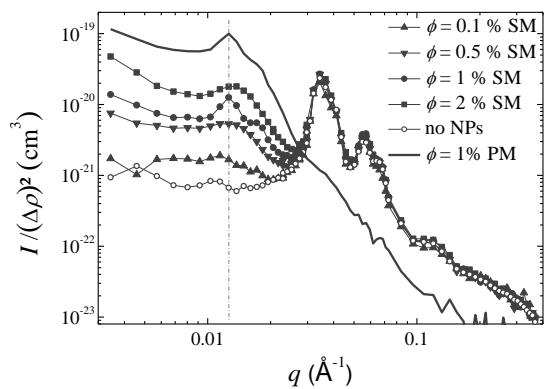

**Fig 2:** Scattering intensity versus wave vector for silica-matched (SM) samples with various amounts of silica nanoparticles, $\phi$ (closed symbols), for a sample without nanoparticles (open symbols) and for a polymer-matched (PM) sample with $\phi=1\%$ (line). Data are normalized by $\Delta\rho^2$, where $\Delta\rho$ is the contrast between the polymer and the solvent for SM samples and between silica and solvent for the PM sample. All samples are prepared with a temperature rate $\dot{T}=0.007°C/min$. The dotted line indicates the position of the silica structure peak.

Contrary to the intermediate and high-$q$ regime, the low-$q$ part of the scattering curves shows a strong evolution with addition of silica, as the normalized intensity continuously increases with nanoparticle volume fraction. An increase larger than a decade is reached for $\phi=2\%$ at very low $q$-vectors. Moreover, the scattered intensity in the low-$q$ region is not a monotonically decreasing function of the wave vector, but instead exhibits a peak around 0.0125 Å$^{-1}$. The peak is visible for all samples except the sample with the lowest NP concentration (0.1%); it is very broad for $\phi=0.5\%$ and 2% but is surprisingly much sharper for $\phi=1\%$. Interestingly, the peak for a silica-matched sample with 1% volume fraction of silica NPs has a similar shape and is located at the same position as the scattering peak of a polymer-matched sample, with 1% volume fraction of silica NPs, where only the silica is visible (Fig. 2). The mapping between the low-$q$ peak in the silica-matched and polymer-matched samples, and the fact that the intensity at low $q$s in silica-matched samples increases monotonically with the volume fraction of silica NPs, suggest that the features of the scattered intensity at low wave-vectors are strongly influenced by the nanoparticles.

At first sight, the influence of the silica NPs on the scattering is rather surprising, since the silica nanoparticles have been carefully matched, as checked for silica NPs alone in the appropriate solvent mixture. Thus, the silica NPs are here completely invisible to neutron scattering. Due to the amount of available micelle-free space, "hole-scattering" based on the fact that polymers are excluded from the volume occupied by the silica NPs is insufficient to explain the large signal observed at low $q$ (see the Supplementary Information). Therefore, the polymer molecules must have been rearranged in the vicinity of the nanoparticles.

The rearrangement of the micelles around the nanoparticles may be of various kinds; one which can be rationalized particularly easily is based on the idea that some polymer molecules may adsorb on the silica surface. The nanoparticles would then be surrounded by a polymer layer, which gives a scattering contrast to the entire object. Indeed, Pluronics molecules are known to adsorb on silica [35, 36], with typical quantities of 0.25-0.50 mg of Pluronics F108 per m$^2$ of silica surface area. To quantitatively check this hypothesis, we convert the low-$q$ intensity increase with respect to the NP-free sample in a mass of adsorbed polymer, knowing both the scattering contrast between the adsorbed layer (polymer-solvent), and the number density of silica particles. If $I_{low}$ is the scattered intensity at low $q$s for the samples with silica nanoparticles, the excess $E$ of normalized low-$q$ intensity with respect to a sample without silica is defined by

$$E = \frac{I_{low}}{(\Delta\rho)^2_{pol-sil}} - \frac{I^0_{low}}{(\Delta\rho)^2_{pol-D_2O}} \tag{1}$$

where $I^0_{low}$ is the intensity for the sample without nanoparticles, and $(\Delta\rho)_{pol-sil} = \rho_{pol} - \rho_{sil}$ is the contrast between the polymer and the silica, and $(\Delta\rho)_{pol-D_2O} = \rho_{pol} - \rho_{D_2O}$ is the contrast between the polymer and D$_2$O. The volume $V_{layer}$ of the polymer layer around one silica nanoparticle reads

$$V_{layer} = \sqrt{\frac{E}{\phi} V_{sil}} \tag{2}$$

where $V_{sil}$ is the average volume of one silica nanoparticle and $\phi$ is the silica volume fraction. The mass of polymer "adsorbed" per unit surface of silica, $\Sigma_{layer}$, is derived from a simple geometric argument:

$$\Sigma_{layer} = \frac{V_{layer}}{S_{sil}} \times d_{pol} \tag{3}$$

where $S_{sil}$ is the average surface of one silica particle, and $d_{pol}=1$ g/ml is the estimated dry density of the polymer.

Experimentally, we take for the intensities at low scattering vector, $I_{low}^0$ and $I_{low}$, the data averaged in the $q$-range where the scattered intensity is rather flat (between 0.0069 Å$^{-1}$ and 0.0103 Å$^{-1}$). The resulting adsorbed masses per unit particle surface, listed in Table 1, are roughly constant. Note that the same analysis has been carried out for silica-matched samples incorporating various amounts of smaller silica nanoparticles (diameter $d=19$ nm), yielding similar numerical values for the measured adsorbed mass unit particle surface (Table 1). We find an adsorbed mass per unit particle surface of $(1.09 \pm 0.16)$ mg/m$^2$, independently of the size and volume fraction of the nanoparticles in the samples, as expected for a quantity which should only depend on the total surface area of silica. Moreover, the numerical estimated values are of the same order of magnitude as those (0.25-0.50 mg/m$^2$) found in the literature for silica nanoparticles [35] or silica macroscopic surfaces [36], although somehow larger. To summarize, the visibility of silica nanoparticles in contrast-matched situations is compatible with the formation of an adsorbed layer of macromolecules on the silica surface.

| $d$ [nm] | $\phi$ | $\dfrac{I_{low}}{(\Delta\rho)^2_{pol-sil}}$ [cm$^3$] | $\Sigma_{layer}$ [mg/m$^2$] |
| --- | --- | --- | --- |
| 19 | 0 | 7.66 10$^{-22}$ | -- |
| 19 | 0.001 | 1.14 10$^{-21}$ | 1.02 |
| 19 | 0.005 | 2.25 10$^{-21}$ | 0.91 |
| 19 | 0.01 | 3.8 10$^{-21}$ | 0.92 |
| 19 | 0.02 | 1.32 10$^{-20}$ | 1.32 |
| 30 | 0 | 7.45 10$^{-22}$ | -- |
| 30 | 0.001 | 1.69 10$^{-21}$ | 1.29 |
| 30 | 0.005 | 4.73 10$^{-21}$ | 1.18 |
| 30 | 0.01 | 6.52 10$^{-21}$ | 1.01 |
| 30 | 0.02 | 1.41 10$^{-20}$ | 1.08 |

**Table 1:** Amount of adsorbed polymer per unit silica surface for two sizes of nanoparticles, diameter $d = 30$ nm

and $d = 19$ nm. This quantity has been evaluated from the intensity at low-$q$ (see text for details), for silica-matched samples incorporated at different volume fraction $\phi$ of silica and prepared with a temperature rate of 0.007°C/min, resp. 0.02°C/min, for particles with $d = 30$ nm, resp. $d = 19$ nm.

## 3.2. Local structure of silica nanoparticles dispersed in the micellar crystals

The scattering of the silica nanoparticles of a series of polymer-matched samples has been measured by contrast-variation SANS, and the scattered intensity normalized by the silica volume fraction is shown in Fig. 3. At large scattering vectors, where only the features of individual particles are probed, all scattering curves collapse, as expected since the nanoparticles are identical for all samples. An estimate of the NP size can be deduced from the Porod domain where one expects $I = Aq^{-4}$, as indeed observed in our data. The proportionality constant is $A = 12\pi \times \dfrac{(\Delta\rho)^2 \phi}{d}$, where $\Delta\rho$ is the contrast between silica and the solvent. We find a diameter of $(25.1 \pm 0.7)$ nm, a numerical value consistent with the diameter of $(30 \pm 6)$ nm measured by transmission electron microscopy. At low $q$, the intensity increases when $q$ decreases, which indicates large-scale (of the order of 100 nm at least) density fluctuations. We find moreover that the intensity at very low $q$ increases monotonically with the heating rate. For all samples prepared with heating rates of 0.001 °C/min, 0.007 °C/min and 0.02 °C/min and NP volume fractions of 0.5%, 1% and 2%, a peak is visible at intermediate wave vectors (see Fig. 3b for a zoom of the data at pH=9 and $\phi = 1\%$ in this $q$ region). A very weak peak is also detected for $\dot{T} = 2$°C/min and $\phi = 1\%$.

Before discussing further the peak, we note that the scattering curves for samples prepared at $\dot{T} = 0.007$°C/min and $\phi = 1\%$ but at lower pH (pH = 5 and 7, respectively) are essentially indistinguishable from the data for the corresponding sample with pH = 9. Decreasing the pH from 9 towards the isoelectric point of 2 is known to reduce the charge on the silica NPs [37, 38], thereby potentially modifying the silica-silica interaction potential. The fact that no significant change in the NPs structure is observed when reducing the pH suggests that the presence of the copolymer matrix and of the copolymer adsorbed on the silica is more important than direct electrostatic interactions in shaping the effective interparticle interactions.

We now turn back to the peak observed in the low-$q$ region. Its position, $q_{peak} \sim 0.0125$ Å$^{-1}$, is rather insensitive to the sample preparation and composition (Fig. 4a), varying by at most 15% when the particle volume fraction varies by a factor of 4, and the heating rate by more than 3 orders of magnitude.

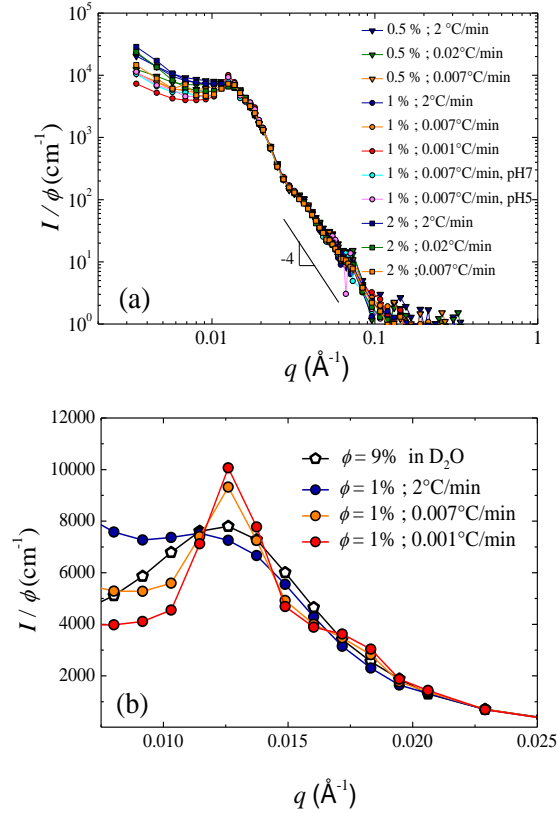

**Fig 3:** (a) Scattered intensities for various polymer-matched samples, at various pH (9, 5 and 7) with different volume fractions of nanoparticles (0.5%, 1% and 2%), and prepared with different temperature ramps (0.001 °C/min, 0.007 °C/min, 0.02 °C/min and 2 °C/min), as indicated in the legend. When not mentioned, pH is 9; (b) Zoom of the structure peak (filled symbols) for samples with $\phi = 1\%$ prepared with 3 different ramps, as indicated in the legend, and (open symbols) for silica NPs at volume fraction 9% dispersed in pure D$_2$O.

A peak in the structure of rather dilute suspensions of silica nanoparticles is usually attributed to long-range repulsive interactions between the nanoparticles, whose concentration controls the position of the peak. One can evaluate the volume fraction of silica nanoparticles from the peak position by using a simple geometric model based on the assumption that, on average, the particles fill the space as in a cubic lattice:

$$\phi = V_{sil} \times (q_{peak}/2\pi)^3 \qquad (4)$$

, where $V_{sil}$ is the average volume of a single nanoparticle. For our samples, the silica volume fraction estimated from this simple model is about 15%. Surprisingly, this value is much larger than the actual volume fraction used to prepare the samples (between 0.5 and 2%). This discrepancy is incompatible with the hypothesis that the silica particles are homogeneously dispersed in the copolymer matrix. It rather suggests that regions where the local volume fraction of NPs is much higher than average coexist with regions where the NPs are diluted, and that the former are responsible for the peak. To better quantify the volume fraction of the concentrated regions, we compare the peak position in the NP-copolymer samples to that measured for suspensions of the same silica NPs but in pure D$_2$O. For the latter, the resulting $q_{peak}$–values vary with the silica volume fraction as a power law with an exponent 1/3, as expected from Eq. (4) (stars in Fig. 4a). The best fit of the data with $q_{peak} = B\,\phi^{1/3}$ gives for the fitting parameter $B = (0.280 \pm 0.004)$ nm$^{-1}$, yielding an average diameter of the silica nanoparticles of $(27.8 \pm 0.4)$ nm, in very good agreement with the diameter of $(30 \pm 6)$ nm measured by transmission electron microscopy. The local silica volume fraction of the concentrated regions can then be evaluated from a comparison between the peak position in the NP-copolymer suspensions, in either polymer- or silica matching conditions, and the peak position in suspensions of nanoparticles in pure D$_2$O (Fig. 4a). We find the local NP volume fraction to range between 7% and 10%, a value considerably larger than the average volume fraction of silica in the sample, thus confirming that silica particles are not distributed homogeneously in space. One may therefore picture the sample as composed of zones of high and low density of nanoparticles, the low-density zones being sufficiently dilute not to present a silica structure factor peak, and the structure peak originating from the high-density zones. We will discuss the nature of the low- and high-density zones in the next section.

We finally note that, although at the same position, the peak of the polymer-matched sample can be significantly sharper than that of a sample composed uniquely of silica nanoparticles in $D_2O$ at 9% volume fraction. This is reminiscent of what observed in experiments where silica nanoparticles suspended in water locally organized in a denser arrangement upon freezing the water: in Ref. [39] it was indeed shown that the packing of the nanoparticles induced by freezing was different from that obtained with nanoparticles in liquid water. Our observations might also originate from a change of the interactions between the silica nanoparticles in the presence of the polymer, due to the layer of polymer adsorbed on the particles. This would increase the repulsion at short distances, possibly leading to a sharper peak.

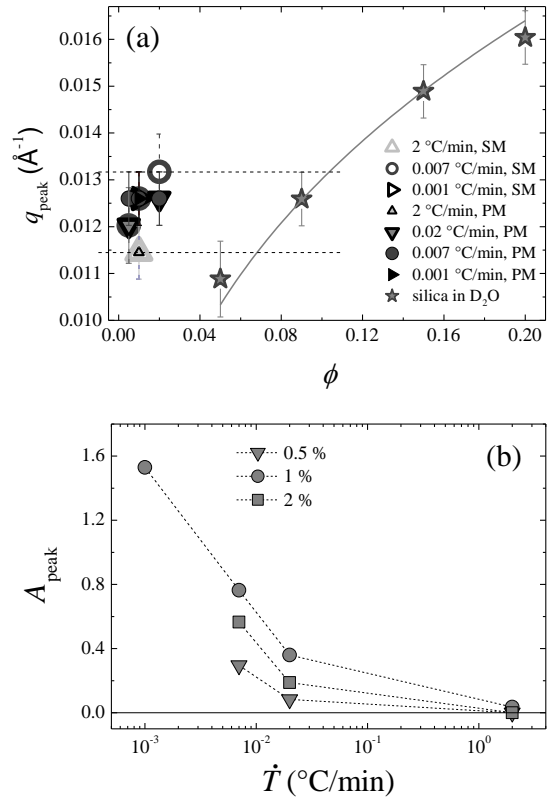

**Fig 4:** (a) Peak position of silica nanoparticles as a function of the nominal silica volume fraction. Data are extracted from the scattering of polymer-matched (PM) and silica-matched (SM) samples, prepared at different temperature rates, as indicated in the legend. Peak positions are compared to those of pure silica suspensions in $D_2O$ at different volume fractions. The line is a fit with a power law with an exponent 1/3,

see Eq. (4). (b) Amplitude of the silica structure peak, as extracted from the scattering of polymer-matched samples (fig. 3), as a function of the temperature rate, for samples with $\phi = 0.5$, 1 and 2%.

It is unclear whether there is any systematic evolution of the peak position with the NP volume fraction and heating rate, $\dot{T}$. If there is any, it must be a very weak dependence, hindered by the low $q$-resolution of the neutron scattering data. By contrast, the amplitude of the peak depends strongly on $\dot{T}$. We define the amplitude of the peak relative to the background as $A_{peak} = (I_{peak} - I_{min})/I_{min}$, where $I_{peak}$ is the scattered intensity measured at the peak position, $q_{peak}$, and $I_{min}$ is the minimum scattered intensity measured at $q$ lower than $q_{peak}$. We have plotted in Fig. 4b the results obtained for NP volume fractions of 0.5%, 1% and 2%. In all cases, we find the amplitude of the structure peak for silica to decrease steadily as the heating rate increases. For $\phi = 1\%$, we find that the amplitude almost vanishes for the fastest temperature ramp, while no peak is detected at this temperature ramp for samples with $\phi = 0.5$ and 2%.

### 3.3. Structure of the regions where silica nanoparticles are more concentrated than average

Having established that the NP-copolymer suspensions contain regions where the silica concentration is larger than average, the crucial question is how these regions are distributed spatially. The significant upturn of $I(q)$ for $q \rightarrow 0$ seen in the neutron data of Fig. 4a suggests that the concentration fluctuations of NP are likely to extend beyond the length scales accessible to SANS. In order to address quantitatively this issue, we perform standard, small-angle and ultra-small angle static light scattering measurements on a sample with $\phi = 0.5\%$ and prepared with a heating rate of 0.02°C/min. We have checked that the light scattering signal is dominated by the scattering from the NPs (compare the symbols to the black line in Fig. 5). Thus, light scattering data directly probe the structure of the silica particles, in analogy to the polymer-matched SANS measurements. As seen in Fig. 5, the light scattering and SANS data collectively span more than four decades in wave-vector and clearly show that the upturn detected at low $q$ in SANS extends over a broader range of $q$. Indeed, a powerlaw variation of $I(q)$ with an exponent close to -3 is measured for $q$ in the range $3 \times 10^{-4}$ - $2 \times 10^{-3}$ Å$^{-1}$. At even lower scattering vectors, a broad peak is seen, whose position, $q_0 \approx 4 \times 10^{-5}$ Å$^{-1}$, corresponds to a length scale on the order of $2\pi/q_0 \approx 16$ µm. Importantly, light scattering data collected for a NP-copolymer suspension with the same amount of NPs ($\phi = 0.5\%$), but a slightly lower copolymer concentration (26% w/w), so that the sample is fluid and not crystallized at room temperature and the NPs are uniformly dispersed, show a much lower and featureless intensity level (gray line in Fig. 5). This confirms that the light scattering signal is due to the regions where silica particles are more concentrated than average.

The presence of NP-rich regions in a composite material was observed recently by small-angle X-Ray scattering (SAXS) [13], where a low-$q$ peak in the SAXS region similar to the low-$q$ peak in our SANS data was measured. These regions were identified with particle-rich grain boundaries. Moreover, a low-$q$ peak similar to that indicated by the arrow in Fig. 5 in the USALS/SALS regions has been observed recently in USALS measurements on a colloidal polycrystal [40], where it was attributed to scattering from the grain boundaries between crystallites. To check whether the peak may have the same origin in our system, we show in the inset of Fig. 5 a microscope image of the same sample as for the scattering data. A network of grain boundaries is clearly visible, whose characteristic size is ~14 µm, very close to that obtained from the position of the low-$q$ peak in the light scattering data. By scanning the focal plane of the microscope through the sample, we check that the dark lines visible in the inset of Fig. 5 are indeed two-dimensional sections of a three-dimensional, foam-like network of closed surfaces, as expected for grain boundaries. Moreover, control experiments on samples that contain the same amount of NPs but are in the fluid micellar phase, and on micellar crystals with no silica particles exhibit uniformly grey microscope images (data not shown). This demonstrates on the one hand that silica NPs are not visible by microscopy when they are uniformly dispersed (as in the fluid sample), and on the other hand that grain boundaries not decorated by nanoparticles are not directly visible (as in the NPs-free micellar crystals). Thus, the dark lines in the inset of Fig. 5 must correspond to the silica-rich regions responsible for the light scattering and the low-$q$ SANS signal in the main panel.

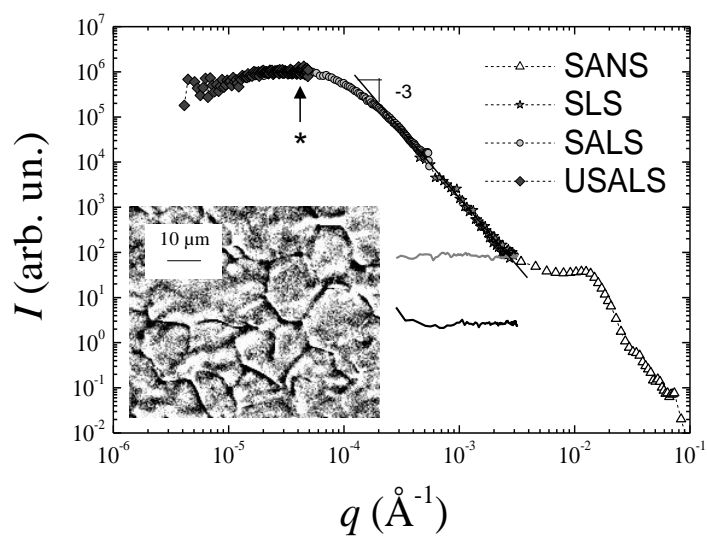

**Fig 5:** Scattered intensity versus wave-vector for a sample with $\phi = 0.5\%$ prepared with a heating rate of 0.02°C/min. The data have been obtained using small-angle neutron scattering (SANS, white triangles), static light scattering (SLS, gray stars), small-angle light scattering (SALS, light gray circles) and ultra-small-angle light scattering (USALS, dark gray diamonds). SANS intensities have been multiplied by an arbitrary factor in order to match the SLS data. The continuous line is a power law fit yielding an exponent of -3. The arrow and the * symbol point to the position of the broad peak at small wave vectors. Gray line: light scattering data for a fluid sample (copolymer concentration 26% w/w) with the same amount of NPs ($\phi = 0.5\%$). Black line: light scattering data for a NP-free micellar crystal. Inset: microscope image of the same sample as for the symbols in the main panel, taken with an x63 objective and a DIC illumination. The network of grain boundaries is clearly visible.

## 4. Discussion

The first important finding of the experiments reported here is that the crystalline face-centered cubic order of the block-copolymer micelles is very robust, as it is not perturbed by the various crystallization rates imposed to the sample or by the presence of moderate amounts of silica nanoparticles (up to 2% v/v). This latter result is consistent with those found by others for mixtures of Pluronics and nanoparticles of various kinds, when the amount of nanoparticles is sufficiently low [20, 41]. Note however that here the nanoparticles and the copolymer micelles have the same spherical shape, which was not the case in Refs. [20, 41]. Based on previous work on colloidal alloys of particles with the same shape and comparable size [22, 23], one might have expected a modification of the unit cell.

A second, unexpected, finding is the presence of a peak in the SANS low-$q$ regime, which is observed when the silica nanoparticles are dispersed in the polymer crystalline phase, whereas no peak is measured when the same amount of silica nanoparticles is dispersed in $D_2O$. We have interpreted this structural peak as the signature of regions of the samples enriched in nanoparticles by a factor up to 10. Light scattering and microscopy measurements reveal that these silica-rich regions can be identified with the texture of grain boundaries that separate micellar crystallites. An analogy can thus be drawn between our samples and atomic or molecular alloys, where impurities are known to segregate in the network of grain boundaries [42, 43]. Following this analogy, one may expect that the segregation process be influenced by both the rate at which the sample is crystallized and the amount of nanoparticle impurities added to the matrix. Figure 6 shows microscopy images taken for samples prepared under various conditions: the differences in the grain boundaries network are evident and support this notion. A systematic study of the effect of $\phi$ and $\dot{T}$ on the grain boundaries texture goes beyond the scope of this work, and is addressed in Ref. [24], where a confocal microscopy investigation of samples doped with fluorescently labeled polystyrene nanoparticles is presented.

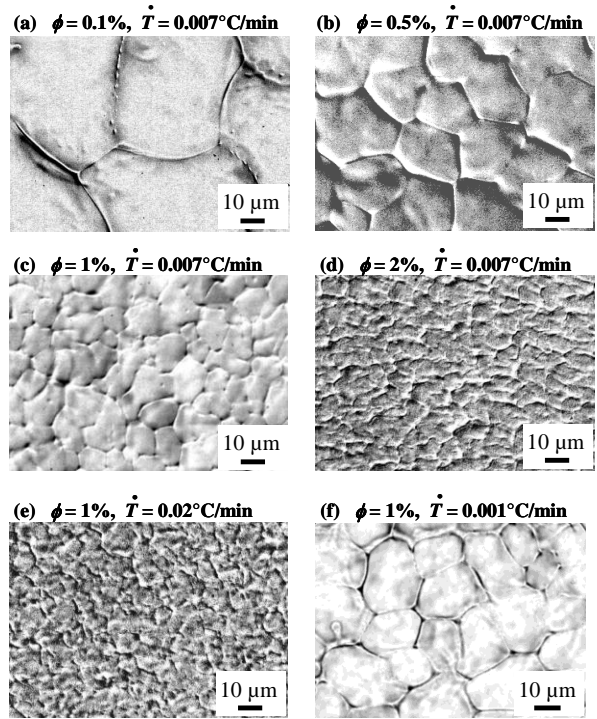

**Fig 6:** Optical microscopy images of several samples prepared at fixed $\dot{T}$ = 0.007 °C/min and various nanoparticle content (a-d), and fixed $\phi$ = 1% and variable heating rates (c,e,f), as indicated in the legend. All images are obtained using differential interference contrast, except image (f) which is obtained using bright field.

A remarkable finding of the neutron scattering data presented here is the fact that the concentration of the silica NPs in the grain boundaries, as inferred by the SANS low-$q$ peak position, seems to be roughly constant in the range 7%-10%, regardless of sample composition and preparation. Note that, even if we take into account an adsorbed polymer layer of average thickness 3 nm [35] which may cover the silica NPs, the effective volume fraction of nanoparticles in the grain boundaries is in the range (12%-17%), hence still much smaller than the volume fraction for a (random) close packing of spheres. This indicates that the NPs experience effective repulsive interactions, which may originate from their charge and from the adsorption of copolymer on the silica surface, as well as from the presence of micelles in the grain boundaries. Since no change in the silica structure is measured when varying the NP surface charge by reducing the sample pH, electrostatic repulsion is likely to play a minor role in the interparticle interactions, at least at the modest silica concentrations that we have tested.

The amplitude of the silica peak exhibits some intriguing features: it decreases when $\dot{T}$ increases and is non-monotonic with respect to $\phi$ (Fig. 4b). Concerning the $\dot{T}$ dependence, the simplest interpretation would be that the volume occupied by the silica-rich regions decreases when increasing the heating rate. Microscopy observations (see Fig. 6c, e and f) show that the grain size decreases as $\dot{T}$ increases, implying that the grain boundary surface per unit volume is larger for the fastest ramps. If the volume occupied by the silica-rich regions was indeed smaller for the faster heating rates, the thickness of the grains should then decrease with $\dot{T}$, in order to compensate for the larger amount of grain boundary surface. Unfortunately, it is difficult to test directly this hypothesis, because microscopy images lack the resolution required to measure precisely the grain boundary thickness. An alternative picture would be that the increase of the peak amplitude be the signature of a higher degree of order of the nanoparticles in the grain boundaries, in analogy to what reported for impurities in freezing water suspensions [39].

Concerning the dependence of the silica structure peak on NP concentration, we find that the peak is systematically more pronounced for samples with a silica volume fraction of 1%, as compared to the peak for $\phi = 0.5$ % and 2% (Fig. 4b). Because the peak position hardly changes with $\phi$ (see Figs. 3b and 4a), one has to rule out that the pronounced peak observed for $\phi = 1\%$ is due to a larger local volume fraction of NP in the silica-rich zones, as compared to the samples with $\phi = 0.5\%$ or $\phi = 2\%$. The change in peak height must then be related either to the nature of the structuring of the NPs in these regions, in analogy to what discussed for the $\dot{T}$ dependence of the peak, or to the $\phi$-dependence of the volume fraction occupied by the silica-rich regions. The volume fraction occupied by the silica-rich regions scales as $t/R$, where $t$ is the thickness of the grain-boundaries and $R$ is the average size of the grain. Microscopy observations suggest that the grain size $R$ decreases monotonically with increasing $\phi$ (see Fig. 6 a-d). A decrease of the grain-boundary thickness $t$ could then explain the evolution of the peak with NP content. Unfortunately, as already mentioned, microscope measurements lack the resolution required to address quantitatively this issue.

As a final remark, we note that in the experiments by Walker and coworkers [33, 41, 45] on a similar particle-doped Pluronic copolymer where the NPs had a size comparable to that of the micelles, as in our case, no structuring of the silica nanoparticles was detected. The reason for this discrepancy is likely to lie in the heating rate used to crystallize the samples. Although the temperature history of the samples investigated by SANS in Refs. [33, 41, 45] is not explicitly mentioned, it is reasonable to assume that extremely slow temperature ramps were not imposed. Our experiments show that ultra-slow ramps, lasting several hours, are in fact required to expel efficiently the nanoparticles from the crystalline grains and confine them in the grain boundaries.

## 5. Conclusion

Using neutron scattering, light scattering and microscopy, we have studied the structure of a composite NP-copolymer material, from the microscopic scale of the micelles constituting the copolymer crystalline matrix up to the mesoscopic scale of the grain boundary texture.

We have found that nanoparticles of size comparable to that of the micelles do not disrupt the local crystalline order, but rather tend to concentrate in the grain boundaries between crystallites. The characteristic size of the grain boundaries network depends on both the crystallization rate and the nanoparticle concentration. The results presented here underline the fact, that, in general, colloidal particles dispersed in a crystalline or liquid crystalline matrix are not uniformly dispersed in the structured matrix, but may partly be confined in specific defects of the matrix. This should be kept in mind when using the colloids as probes of the structure and dynamics of the matrix.

Measurements performed at different pH values suggest that the phenomena observed here are robust with respect to a change in the interparticle interactions. Indeed, we have observed a similar mesoscopic structuring also in samples loaded with polystyrene particles of various kinds and comparable size [24]. The processes described in this paper are thus generic in that they do not rely on the specific surface chemistry of the nanoparticles, and should hold for a large variety of hydrophilic nanoparticles. In particular, the use of suitable NPs may confer specific (optical, magnetic, semi-conducting) properties to this class of soft composite materials, opening routes to the design of original materials structured on the scale of tens of microns. On the other hand, the optical transparency of the matrix allows one to use optical microscopy or light scattering to characterize the mesoscopic grain boundary network, e.g. as a function of sample composition and preparation protocol or under an external load, thus allowing one to address fundamental questions on the crystallization kinetics and mechanical properties of polycrystalline materials.

**Acknowledgements:**

We thank F. Cousin (LLB) for help with the small angle neutron scattering measurements. This work has been supported by ANR under Contract No. ANR-09-BLAN-0198 (COMET).

**Supporting Information Available**: Evaluation of hole scattering for a sample made of invisible silica nanoparticles dispersed in a micellar polycrystal.

# Supporting Information

**Estimation of hole scattering due to matched silica beads in dense micellar assemblies using Monte Carlo simulations**

The presence of the invisible silica nanoparticles (NPs) in dense micellar solutions or crystals may impact the scattering pattern due to the excluded volume effect of the NPs. The goal of the simulations presented below is to check whether this effect can explain our experimental observation (figure 2 of the manuscript) that at low wave vector the scattered intensity increases continuously with the NPs concentration, and can be more than 10 times higher with a NP volume fraction of 2% than without NPs, eventually exhibiting a peak.

The intermicellar pair correlation function, the Fourier transform of which is measured in a silica-matched experiment, is modified due to the presence of holes created by the NPs in the micellar structure. Here we evaluate the importance of this effect on the small angle scattering for a sample containing a volume fraction of 9% of silica nanoparticles. The choice of the volume fraction is motivated by the fact that, although the average NPs volume fraction is never higher than 2%, in the grain-boundaries the NPs volume fraction is higher, typically in the range 7%-10%. The exact structure of the micellar crystal is not of importance here - as long as it is of density comparable to the experimental one, because the increased scattered intensity occurs for wave vectors smaller than the first Bragg peak due to the crystalline order of the micelles.

The simulation proceeds in three steps. First, the structure of the suspension of silica NPs is described using a reverse Monte Carlo algorithm. The analysis is based on the experimental scattering of a (micelle-free) 9% solution of silica nanoparticles (NP diameter $d = 30$ nm, polydispersity $\sigma = 0.20$, log-normal distribution of radii) in water. 604 NPs are arranged in a box of dimension L= 484.4 nm, respecting the excluded volume condition, such that the silica-silica structure factor reproduces the experimentally observed one. Details on such simulations can be found in ref. [S1]. In Fig. S1, a typical snapshot of a resulting silica structure is shown. In Fig. S2, the corresponding structure factor is compared to the experimental one.

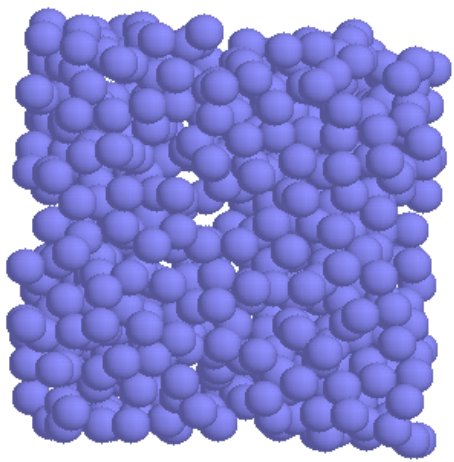 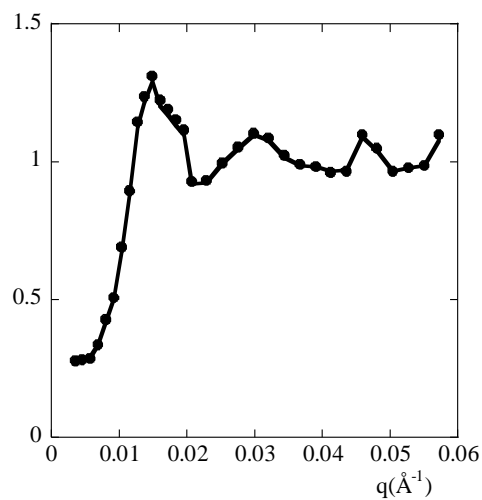

**Figure S1:** Typical snapshot of the structure of the silica beads compatible with the measured structure factor $S(q)$. Note that the diameter of the beads is not drawn to scale

**Figure S2:** Comparison between measured $S(q)$ (symbols) and simulated average structure factor of silica nanoparticles (solid line).

The second step is to create a dense micellar solution (micelle diameter = 17.6 nm, volume fraction of micelles 34%) by placing 13500 micelles (=15x15x15 unit cells containing 4 micelles each) on a face-centered cubic (FCC) lattice in a box with the same dimensions as the one used for the silica NPs. The micelles are then moved around their initial position in a random manner (respecting their excluded volume) for a total of $3x10^6$ time steps to somewhat relax local correlations. The resulting intermicellar structure factor (fig. S3), hereafter called silica-free, shows a first order peak at the same wave vector as the experimental structure factor (0.035 Å$^{-1}$). The second order peak appears at 0.040 Å$^{-1}$, i.e. the local FCC structure of the network is conserved in this ad-hoc model.

The last step consist in copying the silica nanoparticles positioned in step one into the box of micelles created in step two, and suppress all micelles which collide with the NPs. The intermicellar pair correlation is then calculated again. It now contains the hole-scattering contribution.

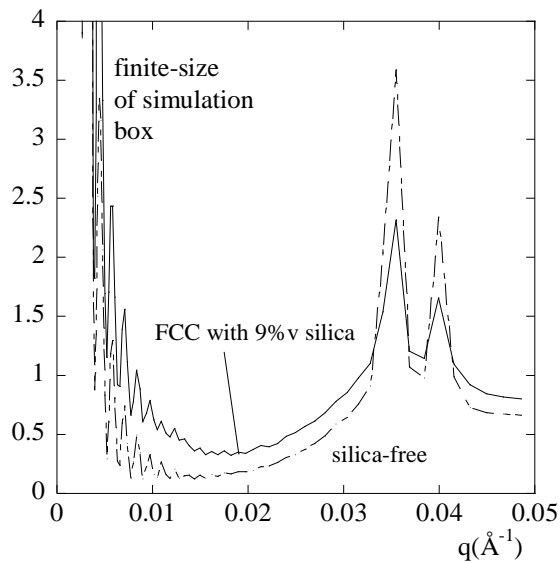

**Figure S3:** Scattering of the original micellar lattice and one where the micelles colliding with (index-matched) silica particles have been removed.

The two structure factors (with and without holes) are compared in Fig. S3. The local FCC structure is shown to be weakened by the holes, but the peak positions are not affected. The low-angle scattering is found to increase by about a factor 2 in roughly the same $q$-range as for the experimental data, but no peak can be detected. This effect is considerably less prominent than the one observed experimentally, where a strong silica peak is visible in the silica-matched case, and where the low-$q$ scattering can be 10 times higher with NPs than without.

As a conclusion, hole scattering cannot account for the experimental observations. Hence, we conclude that a stronger reorganization of the micelles around the silica particles is needed in order to highlight their presence, and polymer adsorption is a credible candidate for such an effect.

[S1] Oberdisse, J.; Hine, P.; Pyckhout-Hintzen, W. *Soft Matter* 2007, *3*, 476